\documentclass[12pt,preprint]{aastex}

\slugcomment{accepted to ApJ}

\shortauthors{Nishikawa et al.}

\shorttitle{Acceleration in relativistic shocks}

\begin{document}

\title{Acceleration Mechanics in Relativistic Shocks by
the Weibel Instability}

\author{K.-I. Nishikawa} 
\affil{National Space Science and Technology Center,
  Huntsville, AL 35805}
\email{ken-ichi.nishikawa@nsstc.nasa.gov}

\author{P. E. Hardee}
\affil{Department of Physics and Astronomy,
  The University of Alabama,
  Tuscaloosa, AL 35487}
\email{phardee@bama.ua.edu}

\author{C. B. Hededal}
\affil{Dark Cosmology Center, Niels Bohr Institute,  Juliane
Maries Vej 30, \\
2100 Copenhagen \O, Denmark}

\and

\author{G. J. Fishman}
\affil{NASA-Marshall Space Flight Center, \\
National Space Science and Technology Center,
  Huntsville, AL 35805}



\begin{abstract}
\vspace{-0.5cm}

Plasma instabilities (e.g., Buneman, Weibel and other two-stream
instabilities) created in collisionless shocks may be  responsible
for particle (electron, positron, and ion) acceleration.  Using a
3-D relativistic electromagnetic particle (REMP) code, we have
investigated long-term particle acceleration associated with
relativistic electron-ion or electron-positron jet fronts
propagating into an unmagnetized ambient electron-ion or
electron-positron plasma. These simulations have been performed with
a longer simulation system than our previous simulations in order to
investigate the nonlinear stage of the Weibel instability and its
particle acceleration mechanism. The current channels generated by
the Weibel instability are surrounded by toroidal magnetic fields
and radial electric fields. This radial electric field is quasi
stationary and accelerates particles which are then deflected by the
magnetic field. Whether particles are accelerated or decelerated
along the jet propagation direction depends on the velocity of
particles and the sign of $\mathbf{E} \times \mathbf{B}$ in the
moving frame of each particle. For the electron-ion case the large
scale current channels generated by the ion Weibel instability lead
to more acceleration near the jet head. Consequently, the
accelerated jet electrons in the electron-ion jet have a significant
hump above a thermal distribution. However, in the electron-positron
case, accelerated jet electrons have more smooth nearly thermal
distribution. In the electron-positron case, initial acceleration
occurs as current channels form and then continues at a much lesser
rate as the current channels and corresponding toroidal magnetic
fields generated by the Weibel instability dissipate.

\end{abstract}

\vspace{-0.5cm}

\vspace{-0.3cm}
\section{Introduction}

\vspace{-0.3cm}

Radiation observed from astrophysical systems containing
relativistic jets and shocks, e.g., active galactic nuclei (AGNs),
gamma-ray bursts (GRBs), and Galactic microquasar systems usually
has nonthermal emission spectra. In most of these systems, the
emission is usually thought to be generated by accelerated electrons
through the synchrotron (jitter) and/or inverse Compton mechanisms.
Radiation from these systems is often observed in the radio through
the gamma-ray region.  Radiation in optical and higher frequencies
typically requires continual strong particle acceleration in order
to counter radiative losses.  It has been proposed that the needed
particle acceleration occurs in shocks produced by differences in
flow speed within the jet (sometimes referred to as ``internal
shocks'') (e.g., Piran 2005).

Particle-in-cell (PIC) simulations can shed light on the
microphysical mechanism of particle acceleration that occurs in the
complicated dynamics within relativistic shocks.  Recent PIC
simulations  of relativistic electron-ion jets show acceleration
occurring within the downstream jet, but not likely by the
scattering of particles back and forth across the shock, as in Fermi
acceleration (Frederiksen et al.\ 2003, 2004; Nishikawa et al.\
2003, 2005; Hededal et al. 2004; Hededal \& Nishikawa 2005; Silva et
al.\ 2003; Jaroschek et al.\ 2005).  In general, these independent
simulations have confirmed that relativistic jets propagating
through a weakly or non magnetized ambient plasma excite the Weibel
instability (Weibel 1959). This instability generates current
filaments and associated magnetic fields (Medvedev \& Loeb 1999;
Brainerd 2000; Pruet et al.\ 2001; Gruzinov 2001), and accelerates
electrons (Silva et al.\ 2003; Frederiksen et al.\ 2003, 2004;
Nishikawa et al.\ 2003, 2005; Hededal et al. 2004; Hededal \&
Nishikawa 2005).

In this paper we present new simulation results of particle
acceleration and magnetic field generation in relativistic jet
shocks using a 3-D relativistic electromagnetic particle-in-cell
(REMP) code. Our new simulations with additional diagnostics allow
us to examine the mechanism of particle acceleration in detail,
e.g., Hededal et al.\ (2004).

At the end of our simulations we see four different regions.  From
the beginning of interaction outwards to the jet front the regions
are: (1) a linear instability growth region, (2) a nonlinear
saturation region, (3) a nonlinear dissipation region, and (4) the
head region of the relativistic jet. The instability generates
current filaments elongated along the streaming direction with
associated transverse toroidal magnetic fields.  The charge
separation  accompanying the current filaments produces radial
electric fields which are mainly perpendicular to the toroidal
transverse magnetic fields.  Particles are accelerated and
decelerated in parallel and perpendicular directions relative to
their initial motion by the transient force due to the radial
electric field and the Lorentz force from the transverse magnetic
field. In \S 2 the simulation model and initial conditions are
described. The simulation results are presented in \S 3, and in \S 4
we summarize and discuss the new results.

\vspace{-0.5cm}
\section{Simulation Setup}

The code used in this study is a modified version of the TRISTAN
code, a relativistic electromagnetic particle (REMP) code (Buneman
1993).  Descriptions of PIC codes are presented in Dawson (1983),
Birdsall \& Langdon (1995), and Hickory \& Eastwood (1988). This
code has been used previously for many applications including
astrophysical plasmas (Zhao et al.\ 1994; Nishikawa et al.\ 1997).

Two simulations were performed using a system with ($N_{\rm x},
N_{\rm y}, N_{\rm z}) = (85, 85, 640)$ cells and a total of  $\sim
380$ million particles (27 particles$/$cell$/$species for the
ambient plasma) in the active grid zones. In the simulations the
electron skin depth, $\lambda_{\rm ce} = c/\omega_{\rm pe} =
9.6\Delta$, where $\omega_{\rm pe} = (4\pi e^{2}n_{\rm e}/m_{\rm
e})^{1/2}$ is the electron plasma frequency and $\Delta$ is the grid
size.  Here the computational domain is two times longer than in our
previous simulations (Nishikawa et al. 2003, 2005).  In both
simulations the electron number density of the jet is $0.741n_{\rm
e}$, where $n_{\rm e}$ is the ambient electron density and as in the
previous work $\gamma  = 5$ (Nishikawa et al.\  2003, 2005). In both
simulations jets are injected in a plane across the computational
grid at  $z = 25\Delta$ in the positive $z$ direction in order to
any eliminate effects associated with the boundary conditions at $z
= z_{\rm \min}$. Radiating boundary conditions were used on the
planes at {\it $z = z_{\min}~{\&}~z_{\max}$}. Periodic boundary
conditions were used on all transverse boundaries (Buneman 1993).
The ambient and jet electron-positron plasma has mass ratio $m_{\rm
e^-}/m_{\rm e^+} = 1$, and the electron-ion plasma ($e^{-}/i^{+}$)
has mass ratio $m_{\rm i}/m_{\rm e} = 20$. The electron/positron
thermal velocity in the ambient plasma is $v^{\rm e}_{\rm th} =
0.1c$ and the ion thermal velocity is $v^{\rm i}_{\rm th} = 0.022c$
where $c$ is the speed of light.

As in previous papers (Nishikawa et al.\ 2003, 2005) the ``flat"
(thick) jet fills the computational domain in the transverse
directions (infinite width). Thus, we are simulating a small section
of a relativistic shock infinite in the transverse direction. The
present choice of parameters and simulations allows comparison with
previous simulations (Silva et al.\ 2003; Frederiksen et al.\ 2003,
2004; Nishikawa et al.\ 2003, 2005; Hededal et al.\ 2004; Hededal \&
Nishikawa 2005). If the simulations are scaled to the interstellar
medium (ISM), the electron skin depth becomes $c/\omega_{\rm pe}
\sim 3 \times 10^{8}~{\rm m}~{\rm s^{-1}} /50,000~{\rm Hz} \sim
6,000~{\rm m}$ and the length of the simulation box is about $4
\times 10^{5}$~m (= 400 km).

\vspace{-0.5cm}
\section{Simulation results}

The jet makes contact with the ambient plasma at a 2D interface
spanning the computational domain.  Here the dynamics of the
propagating jet head and the formation of a shock region is studied.
Effectively we study the evolution of the Weibel instability of a
small portion of a much larger shock in a spatial and temporal way
that includes the motion of the jet head and the spatial development
of nonlinear saturation and dissipation from the injection point to
the jet front defined by the fastest moving jet particles.

Figure 1 shows components of the current density resulting from
development of the Weibel instability behind the jet front at time
$t = 19.5/\omega_{\rm pe}$ [panels (a) $e^{-}/i^{+}$ \& (c)
$e^{-}/e^{+}$] and at time $t = 59.8/\omega_{\rm pe}$ [panels (b)
$e^{-}/i^{+}$ \& (d) $e^{-}/e^{+}$]. The arrows show the $z$ and $x$
components of the current density and the color scale also shows the
$z$ component of the current density. The intensities and arrow
lengths are scaled so that small values can be seen in each panel
and cannot be directly intercompared.
%
%
Panels 1a ($e^{-}/i^{+}$) and 1c ($e^{-}/e^{+}$)  where the $z$
component dominates indicate the initial linear growth of the Weibel
instability. The current channels generated by the Weibel
instability in the electron-positron case show considerable
dissipation at the later time at large distance (panel 1d).  This
result is indicated by the lower values of $j_{\rm z}$ and
disorganized orientation of the arrows, $j_{\rm x,z}$ current
components, for $z/\Delta > 270$.  On the other hand, the current
channel(s) in the electron-ion case remain well organized at the
larger distances at this later time (panel 1b).

Figure 2 indicates magnetic field structure and field energy along the
$z$ direction at time $t = 59.8/\omega_{\rm pe}$. Panels 2a
($e^{-}/i^{+}$) and 2c ($e^{-}/e^{+}$) show 1-D cuts through the
computational grid parallel to the $z$-axis at $x/\Delta = 43$ and
$y/\Delta =$~ 33, 43, and 53.  The spacing in $y$ is about the electron
skin depth ($\lambda_{\rm ce} \sim 9.6\Delta$).
This figure provides some quantitative longitudinal information about
the filament structures shown qualitatively in Figure 1.  With
separation by about the electron skin depth, the phase of the
instability is different along different 1-D cuts, but the amplitudes
are similar.  Panel 2c ($e^{-}/e^{+}$) shows that $B_{\rm x}$ declines
for $z/\Delta > 270$ and dissipates around $z/\Delta = 480$ in the
electron-positron case. This decline coincides with the disorganization
in current structure revealed in Figure 1d.  $B_{\rm x}$ does not
similarly decline in the electron-ion case (panel 2a).
%
%

In order to examine global changes, the magnetic field energies
averaged across the $x - y$ plane are plotted along the $z$
direction in panels 2b ($e^{-}/i^{+}$) and 2d ($e^{-}/e^{+}$).  We
note that very little magnetic energy is in the parallel magnetic
field component. Nonlinear saturation occurs at $z/\Delta = 340$ in
the electron-ion case (panel 2b) and at $z/\Delta = 270$ in the
electron-positron case (panel 2d) (see also Kato 2005).  The
saturation level in the electron-ion case is about two times higher
than in the electron-positron case. While the magnetic field energy
is dissipated around $z/\Delta = 450$ in the electron-positron case,
a much smaller reduction is seen in the electron-ion case.  Here the
ion Weibel instability is excited in the electron-ion case (Hededal
et al.\ 2004) and maintains a large total magnetic field energy
which is more than four times of that in the electron-positron case.
In the electron-positron case the magnetic fields appear turbulent
in the region ($400 < z/\Delta < 500$) as suggested by the current
structure seen in Figure 1d.

The acceleration of electrons by the Weibel instability has been
reported in previous work (Silva et al.\ 2003; Frederiksen et al.\
2003, 2004; Nishikawa et al.\ 2003, 2005; Hededal et al.\ 2004;
Hededal \& Nishikawa 2005).  Some explanation of the acceleration
mechanism associated with the development of current channels can be
found in Hededal et al.\ (2004) and Medvedev et al.\ (2005). Here
Figure 3 shows current channel structure in 2D images in the $x - y$
plane  at $z/\Delta = 430$ and $t = 59.8/\omega_{\rm pe}$, which
correspond to the nonlinear dissipation region. Panels 3a
($e^{-}/i^{+}$) and 3b ($e^{-}/e^{+}$) show the $z$ component of the
current density generated by the Weibel instability in color with
the $x$ and $y$ components of the electric field represented by
arrows.  Panels 3c ($e^{-}/i^{+}$) and 3d ($e^{-}/e^{+}$) show the
$z$ component of $\mathbf{E}\times\mathbf{B}$ along with the $x$ and
$y$ components of the magnetic field shown as arrows. Panels 3a, 3c,
and 3e show that the center of the single ion current channel is
located at $(x/\Delta, y/\Delta) = (35, 72)$. This ion current
channel is surrounded by transverse toroidal magnetic fields
indicated by the arrows in panel 3c. This large scale (nearly the
ion skin depth ($43\Delta$)) current channel is generated by the ion
Weibel instability.  Further investigation of the nonlinear
evolution of the ion Weibel instability requires a larger system
such as $(N_{\rm x}, N_{\rm y}, N_{\rm z}) = (200, 200, 800),$
(e.g., Frederiksen et al.\  2004), that can contain several current
channels in the transverse direction and be sufficiently long to
illustrate the spatial development.
%
%

In addition to toroidal magnetic fields, radial electric fields
accompany the charge separation associated with the current channels
(Milosavljevi\'{c}, Naker, \& Spitkovsky 2005).  These electric
fields are indicated by the arrows in panels 3a ($e^{-}/i^{+}$) and
3b ($e^{-}/e^{+}$). The electric and magnetic fields are
approximately perpendicular in the $x - y$ plane. The $z$ component
of $\mathbf{E} \times \mathbf{B}$ in the simulation/ambient rest
frame is mainly positive (see panels 3c and 3d) on the $x - y$ plane
at this time and indicates the direction of motion of particles
accelerated by the radial electric fields as modified by the Lorentz
force due to the transverse magnetic fields.  It should be noted
that the transient acceleration due to the radial electric field
occurs within one gyromotion.  Therefore, the resultant motion is
not identical to the $\mathbf{E} \times \mathbf{B}$ drift which
takes place over many gyromotions.  Still the sign of the $z$
component of $\mathbf{E} \times \mathbf{B}$ gives an indication of
the average effect of the radial and transverse magnetic fields on
the ambient and jet particles.  Panels 3c and 3d which show that
$(\mathbf{E} \times \mathbf{B})_{\rm z}$ is typically in the $+z$
direction in the ambient reference frame imply that ambient
particles accelerated by the radial electric fields are directed by
Lorentz forces to move in the $+z$ direction.  Thus, on average
ambient particles are accelerated in the $+z$ direction.

The sign of $(\mathbf{E} \times \mathbf{B})_{\rm z}$ depends on the
frame of reference through the Lorentz transformation, e.g., Jackson
(1999), We have computed the value of $(\mathbf{E} \times
\mathbf{B})_{\rm z}$ in reference frames moving with $\beta_z =
v_{\rm z}/ c = 0.6,~0.8~\&~0.9798$ where the third value corresponds
to the initial jet particle velocity.  The results for $(\mathbf{E}
\times \mathbf{B})_{\rm z}$ with $\beta_z = 0.8$ relative to the
simulation frame are shown in Panels 3e and 3f. In panels 3e and 3f
the arrows indicate the $x$ and $y$ components of the magnetic field
in the moving reference frame.  In these reference frames
$(\mathbf{E} \times \mathbf{B})_{\rm z}$ is mainly negative. The
implication is that particles moving with these higher speeds, when
accelerated by the radial electric fields, are on average directed
by Lorentz forces to move in the $-z$ direction in this moving
reference frame. In the ambient reference frame this appears as a
deceleration.  In reality, both spatial and temporal variation in
filament structure produces more than a simple average acceleration
or deceleration along the flow, and more than simple scattering into
the transverse direction.  The complicated nature of the particle
motion and the acceleration process accompanying the filament
structure is illustrated and discussed in Hededal et al.\ (2004).

Figure 4 shows the phase space distribution of jet and ambient
electrons in order to establish how particle acceleration is related
to magnetic and electric field spatial development associated with
the Weibel instability. The value of $(\mathbf{E} \times
\mathbf{B})_{\rm z}$  in a frame moving with $v_{\rm z}/c \sim 0.8$
becomes more strongly negative in the electron-ion case (Fig.\ 3e)
than in electron-positron case (Fig.\ 3f). Consequently, a larger
average deceleration of  jet electrons occurs in the electron-ion
case than in electron-positron case shown in Panels 4a and 4b. In
the electron-ion case, jet electrons are accelerated in the parallel
direction at $z/\Delta =495$ not too far behind the jet head (panel
4a).  The largest perpendicular acceleration occurs at $z/\Delta
=330$ where the magnetic field energy achieves a local maximum (see
Fig.\ 2b). Behavior is qualitatively similar to the
electron-positron case but maximum values are somewhat reduced.
Additionally, the electron-positron case clearly saturates by this
simulation time. Panel 4d clearly shows a linear acceleration region
($130 < z/\Delta \le 270$), a nonlinear region ($270 < z/\Delta \le
540$), and a jet head region ($540 < z/\Delta \le 580$). The
nonlinear region is contained between the two magnetic field energy
peaks at $z/\Delta =$ 270 \& 540 shown in Figure 2d.  By comparison
development in the electron-ion case is affected by the transverse
dimensions of the grid, i.e., one current channel fills the grid
transversely beyond the magnetic energy maximum at $z/\Delta = 330$.
With this restriction removed, it seems likely that even larger
differences would appear between the electron-ion case and
electron-positron case (Hededal et al. 2004).

The parallel acceleration of ambient electrons is shown in panels 4e
($e^{-}/i^{+}$) and 4f ($e^{-}/e^{+}$). At the time $t =
59.8/\omega_{\rm pe}$ the ambient electrons in the electron-ion case
are not accelerated nearly as much as those in the electron-positron
case. It seems likely that ambient ion inertia restricts ambient
electron parallel acceleration, and the ion Weibel instability is
not fully excited (e.g., Hededal et al.\ 2004).  This lack of
parallel acceleration can also be a result of insufficient
transverse grid dimension. By comparison, in the electron-positron
case positrons have the same mass as electrons and the ambient
electrons and positrons are fully involved in the Weibel instability
with our present grid dimensions.

We need to examine the particle momentum distributions in order to
determine whether the instability results in a thermal or a power
law distribution.  Additionally, it is necessary to examine the
particle momentum distributions in the parallel and transverse
directions as particle acceleration need not lead to an isotropic
distribution.  Figure 5 shows momentum distributions of jet
electrons at time $t = 59.8/\omega_{\rm pe}$. Parallel momentum,
$\gamma_v v_{\parallel}$, distributions are shown in panels (a)
$e^{-}/i^{+}$ and (b) $e^{-}/e^{+}$ and perpendicular momentum
distributions, $\gamma_v v_{\perp}$  are shown in panels (c)
$e^{-}/i^{+}$ and (d) $e^{-}/e^{+}$.
%
%
In the panels the blue curves show the distribution of jet electrons
found by plotting one third of the jet electrons in a portion of the
grid near to the inlet ($25 \leqq z/\Delta \lesssim 213$) and
provides a reasonable indication of the initial injected electron
distribution. The red curves show the momentum distribution of jet
electrons found by plotting one third of the jet electrons from a
region behind the jet front ($400 \lesssim z/\Delta \lesssim 590$)
and indicates the accelerated electron distribution. All
distributions are plotted in the ambient/simulation rest frame.

The distribution of a Lorentz-boosted thermal electron population is
also plotted in Figure 5 as dashed green curves. Here we use a
bi-Maxwellian distribution function of the form (see Qin et al.\
2005)
%
$$
f(\gamma_{\rm v}v_{\parallel}, \gamma_{\rm v}v_{\perp}) =
N_{\max}\exp[- (\gamma_{\rm v} m_{\rm e}v_{\parallel} - \gamma_{\rm
u} m_{\rm e} u)^{2}/ (2\gamma_{\rm u}^{3}m_{\rm e} k
T_{\parallel})]\exp[-(\gamma_{\rm v}m_{\rm
e}v_{\perp})^{2}/(2\gamma_{\rm u}m_{\rm e} k T_{\perp}]~,
$$
%
where $v_{\perp}^{2} = v_{\rm x}^{2} +v_{\rm y}^{2}$, $\gamma_{\rm
v} = (1 -(v_{\perp}^{2} +v_{\parallel}^{2})/c^{2})^{-1/2}$ is the
Lorentz factor for electrons, and $\gamma_{\rm u} = (1
-(u/c)^{2})^{-1/2}$ is the bulk (averaged) Lorentz factor.  This
particular form presupposes that the electrons are ``cold'', i.e.,
$m_{\rm e}c^{2}/(kT) > 1$, in the jet rest frame. The parallel and
perpendicular thermal distributions that provide a best fit to the
observed  parallel and perpendicular distributions are calculated
separately but use the same bulk Lorentz factor, $\gamma_{\rm u}$.
The bulk Lorentz factor, different for the electron-ion and
electron-positron cases, is found by fitting the parallel momentum
distributions (red curves in panels (5a) and (5b). Note that the
perpendicular velocity of jet electrons appears as a secondary
effect through $\gamma_{\rm v}$ in the parallel momentum
distributions.  Our best fits require a bulk Lorentz boost of
$\gamma_{\rm u}u \sim$ 3.0 and 3.8 for $e^{-}/i^{+}$ and
$e^{-}/e^{+}$ cases, respectively, i.e., more electron slowing for
the electron-ion case.  The accompanying parallel temperatures are
$T_{\parallel} = 1.16 \times 10^9~K$ and $0.72 \times 10^9~K$ for
$e^{-}/i^{+}$ and $e^{-}/e^{+}$ cases, respectively, i.e, hotter for
the electron-ion case.  The accompanying perpendicular temperatures
are $T_{\perp} = 9.59 \times 10^9~K$ and $4.79 \times 10^9~K$ for
$e^{-}/i^{+}$ and $e^{-}/e^{+}$ cases, respectively, i.e., again
hotter for the electron-ion case.  In both $e^{-}/i^{+}$ and
$e^{-}/e^{+}$ cases $T_{\perp} > T_{\parallel}$. Note that these
fitted temperatures imply a ``warm'' jet electron plasma, i.e., $kT
\sim m_{\rm e} c^2$, and formally the assumption that the
temperatures are cold is violated. Nevertheless, the comparisons
remain qualitatively if not quantitatively robust.

 The electron-ion case shows a significant  hump at $\gamma
v_{\parallel} = 10$, which is found in the parallel velocity
distribution for the electron-ion case (panel 5a).  The
electron-positron case (panel 5b) shows a smoother higher energy
tail without a noticeable hump. The hump could be due to more
effective acceleration near the jet head (near $z\Delta =490$) as
shown in Fig. 4a, also indicated by Fig. 2 in Hededal et al.\
(2004). This more effective acceleration in the electron-ion case is
also indicated by the higher fitted perpendicular temperature in the
electron-ion case (see panels 5c and 5d).

The differences in parallel and perpendicular momentum distributions
between electron-ion and electron-positron cases provides a potential
discriminator between electron-ion and electron-positron shocks.  In
particular, differences in the momentum distributions will appear as
differences in the radiation spectrum (Hededal 2005, Hededal \&
Nordlund 2005).  Such potential spectral differences will be in
addition to more subtle differences induced by differences in magnetic
field strength and structure.

\vspace{-0.5cm}
\section{Summary and Discussion}

We have performed self-consistent, three-dimensional relativistic
particle simulations of relativistic electron-positron and
electron-ion jets propagating into unmagnetized electron-positron
and electron-ion ambient plasmas, respectively.  The present
simulations were performed using a larger simulation system and for
a longer time compared to our previous simulations (Nishikawa et
al.\ 2003, 2005).  The larger grid and longer time allowed us to
investigate the nonlinear development of the Weibel instability and
the accompanying acceleration mechanism. The Weibel instability
forms current filaments and accompanying transverse toroidal
magnetic fields around the current filaments.  Electrons deflected
by the perturbed (small) transverse magnetic fields accompanying the
current filaments subsequently enhance the current filaments (Weibel
1959; Medvedev \& Loeb 1999; Brainerd 2000; Gruzinov 2001). The
acceleration of particles by electric fields toward or away from the
current filaments and deflection of particles due to the Lorentz
force increases as the magnetic field perturbation grows in
amplitude. Basically we see that the transient force due to the
radial electric field with circular magnetic field leads to
acceleration in the perpendicular direction relative to the initial
motion, and an average deceleration in the parallel direction but
with an acceleration of some electrons in the parallel direction to
produce a high energy thermal distribution. It should be noted that
the Weibel instability is convective, i.e., the real part of the
frequency is zero and current filaments grow temporally in the
appropriate rest frame (Medvedev \& Loeb 1999; Nishikawa et al.\
2003, 2005). In our simulations the values of $(\mathbf{E} \times
\mathbf{B})$ fluctuate highly in a spatial sense in both ambient and
jet reference frames. Our results here are consistent with results
from previous simulations (Frederiksen et al.\ 2003, 2004; Hededal
et al.\ 2004; Nishikawa et al.\ 2003, 2005; Hededal \& Nishikawa
2005).

The present simulations show no clear sign of Fermi acceleration and
the main acceleration of electrons takes place in the transition
region behind the jet head. Processes in the relativistic
collisionless shock are dominated by current structures produced by
the Weibel instability.  This instability is first excited near to
the jet injection plane and grows spatially from smaller to larger
current filaments behind the propagating jet front.  Radial electric
fields accompany the development of current channels and toroidal
transverse magnetic fields around the current channels.  The
toroidal magnetic and radial electric fields associated with the
current filaments accelerate and decelerate jet electrons and
positrons. In the electron-positron case the organized current
channels and accompanying fields reach a maximum and then dissipate
and become more  disordered. Particle acceleration/deceleration
effects decrease in the dissipation region behind the jet front.
Here we note that other new simulations (see Nishikawa et al.\
2005a) show that the level of magnetic field in the dissipation
region depends on the initial jet Lorentz factor or the Lorentz
factor distribution. On average bulk jet electron motion is
decelerated in the parallel direction while some jet particles are
accelerated in the parallel direction to form a nearly thermal
distribution ($e^{-}/e^{+}$) or a thermal distribution with a hump
in higher energy ($e^{-}/i^{+}$). Deceleration in the parallel
direction is accompanied by large acceleration in the perpendicular
directions. The ambient electron-positron medium is on average
accelerated in the jet direction in the interaction.  In the
electron-ion case the instability evolves spatially similar to the
electron-positron case but here large scale ion current channels are
ultimately generated by the ion Weibel instability in the
electron-ion jet (see also Frederiksen et al.\ 2004; Hededal et al.\
2004). In this case more bulk deceleration, a prominent humped high
energy tail in the parallel direction, and more transverse
acceleration occur for the jet electrons. Very little acceleration
of the ambient electrons in the electron-ion jet is observed on the
simulation grid. It is clear that a larger simulation grid is
necessary to follow non-linear development in the electron-ion case
for a longer time.

The efficiencies of conversion of bulk kinetic energy into radiation
via synchrotron or ``jitter'' emission from relativistic shocks is
determined by the magnetic field strength and structure, and by the
electron energy distribution behind the shock (Medvedev 2000;
Hededal 2005). Whether the radiation is in the synchrotron or jitter
regime depends only on the nature of the magnetic field. If the
typical fluctuation spatial scale of the magnetic field is smaller
than the expected gyro-orbit, and the magnetic field is turbulent
(power law distributed) then the radiation is in the jitter regime
(Hededal 2005; Hededal \& Nordlund 2005). The radiation spectrum
from the accelerated electrons in the Weibel generated
electromagnetic field shows good agreement with observations (Piran
2005; Hededal 2005, Hededal \& Nordlund 2005).

The fundamental microscopic characteristics of relativistic shocks are
essential for a proper understanding of the prompt gamma-ray and
afterglow emission in gamma-ray bursts, and also to an understanding of
the particle reacceleration processes and emission from the shocked
regions in relativistic AGN jets.  Obviously the accelerated particle
distribution both parallel and perpendicular to the bulk motion needs to be
known and, in particular the generation of magnetic field and its
structure in the shock needs to be known.  Since the shock dynamics is
complex and subtle at the microscopic level, more comprehensive studies
are required to better understand the acceleration of electrons, the
generation of magnetic fields, magnetic field structure and the
associated emission. This further study will provide the insight into
basic relativistic collisionless shock characteristics needed to
provide a firm physical basis for modeling the emission from shocks in
relativistic flows.

\acknowledgments K.I. Nishikawa was a NRC Senior Research Fellow at
NASA Marshall Space Flight Center until June 2005.  K.I. Nishikawa's
research is partially supported by the National Science Foundation
through awards ATM 9730230, ATM-9870072, ATM-0100997, INT-9981508,
and AST-0506719 to the University of Alabama in Huntsville. P.
Hardee acknowledges National Science Foundation support through
award AST-0506666 and the National Space Science \& Technology
Center (NSSTC/MSFC) through award NCC8-256 to The University of
Alabama. The simulations have been performed on IBM p690 (Copper) at
the National Center for Supercomputing Applications (NCSA) which is
supported by the National Science Foundation.

\clearpage

\begin{figure}[ht]
\vspace*{-1.3cm} \epsscale{0.8} \plotone{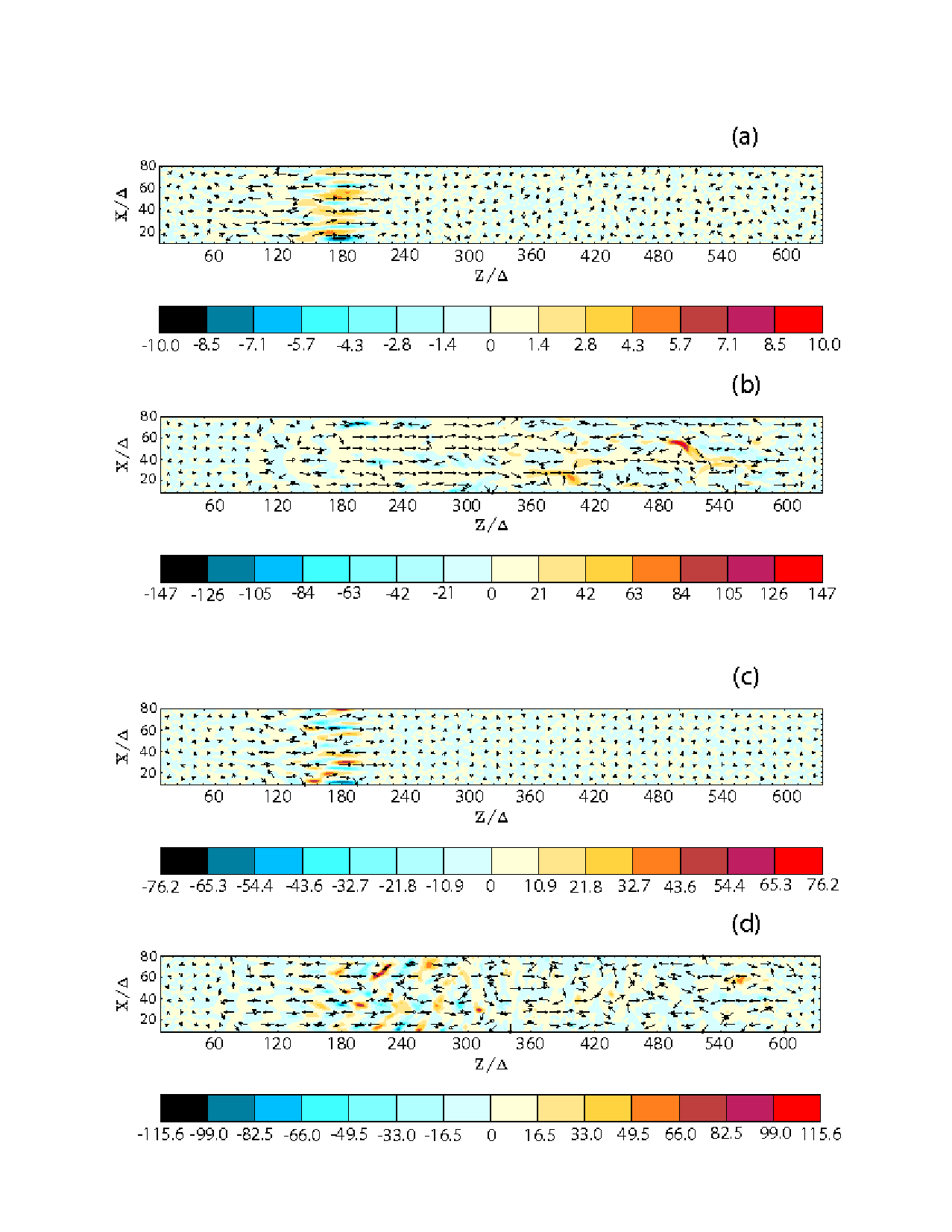}
\vspace*{0.5cm} \caption{2D images in the $x - z$ plane at $y =
43\Delta$ for the electron-ion (panels a \& b) and electron-positron
(panels c \& d) jets injected into an unmagnetized ambient plasma at
$t = $ $19.5/\omega_{\rm pe}$ (panels a \& c), and $59.8/\omega_{\rm
pe}$ (panels b \& d). The colors indicate the $z$ component of
current density generated by the Weibel instability with the $z$ and
$x$ components of current density represented by arrows.}
\end{figure}

\clearpage

\begin{figure}[ht]
\epsscale{0.9} \vspace*{-.5cm} \hspace*{-.0cm} \plotone{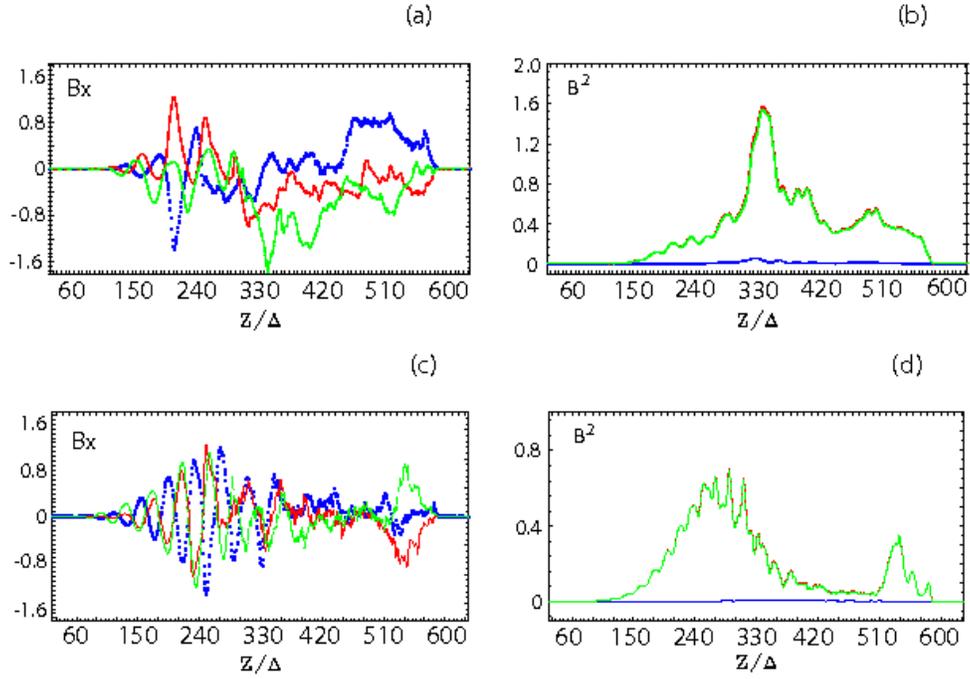}
\vspace*{-.7cm} \caption{1-D cuts along the $z$-direction of the
$x$-component of the magnetic field (panels a \& c), and magnetic
field energies (panels b \& d) at $t = 59.8/\omega_{\rm pe}$ for the
electron-ion (panels a \& b) and electron-positron (panels c \& d)
jets.  1D cuts for (panels a \& c) are at $x/\Delta = 38$ and
$y/\Delta = 33~(blue), 43~(red), 53~(green)$, and cuts are separated
by about an electron skin depth. In panels b \& d the green curves
show the perpendicular magnetic field energy ($B_{\perp} = B_{\rm
x}^{2} +B_{\rm y}^{2}$). Since the parallel magnetic field energy
(blue) is small, the curves of the total magnetic field energy (red)
underlie the perpendicular energy curves (green). The units are
simulation units.}
\end{figure}

\clearpage

\begin{figure}[ht]
\epsscale{0.9} \vspace*{-2.0cm} \hspace*{-0.0cm} \plotone{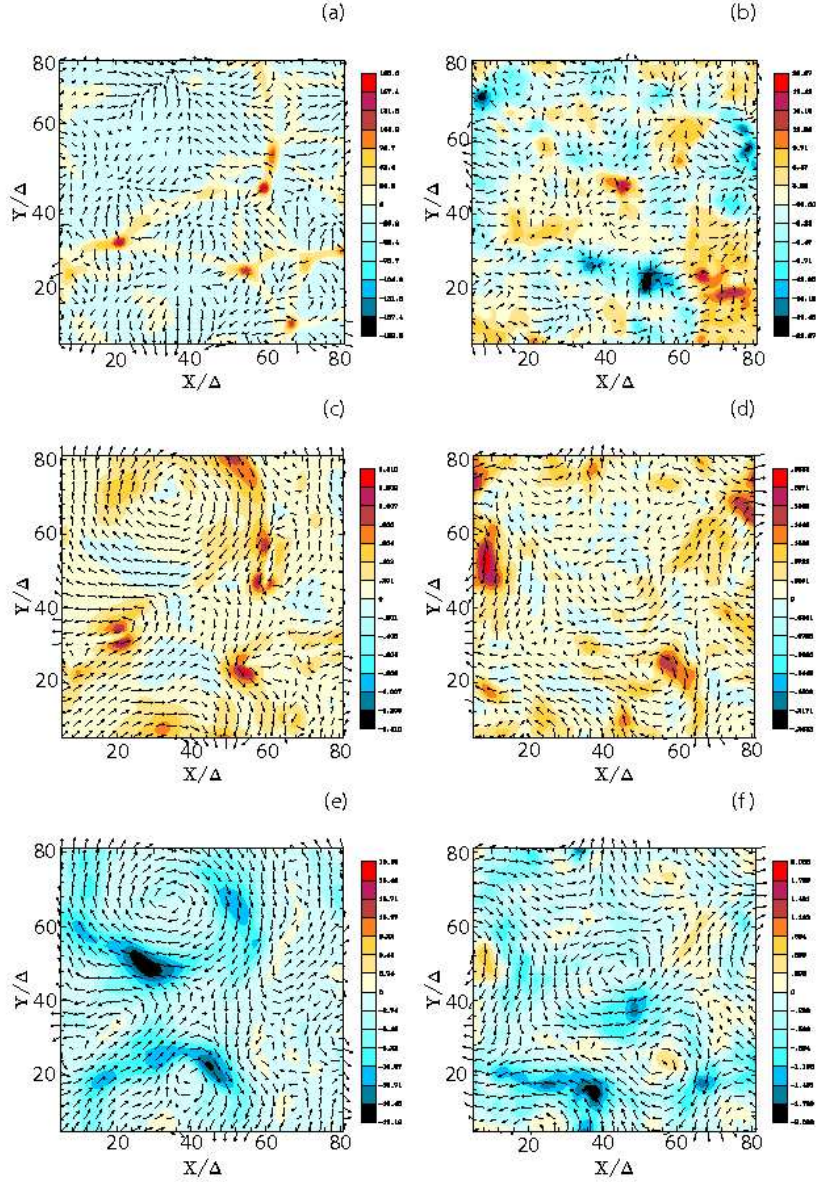}
\vspace*{-2.0cm} \caption{2D images in the $x - y$ plane at
$Z/\Delta = 430$ for the electron-ion case (left column) and
electron-positron case (right column) at $t = 59.8/\omega_{\rm pe}$.
In panels a \& b the colors indicate the $z$ component of current
density with the $x$ and $y$ components of the electric field
represented by arrows. The middle panels (c \& d) show the $z$
component of $\mathbf{E}\times\mathbf{B}$ with the $x$ and $y$
components of magnetic field indicated by the arrows in the grid
rest frame. The bottom panels (e \& f) show the $z$ component of
$\mathbf{E}\times\mathbf{B}$ and  $B_{\rm x,y}$ (by the arrows) in
the jet rest frame  moving with $\beta_{\rm z} = v_{\rm z}/c =
0.8$.}
\end{figure}

\clearpage

\begin{figure}[ht]
\epsscale{0.90} \vspace*{-2.5cm} \plotone{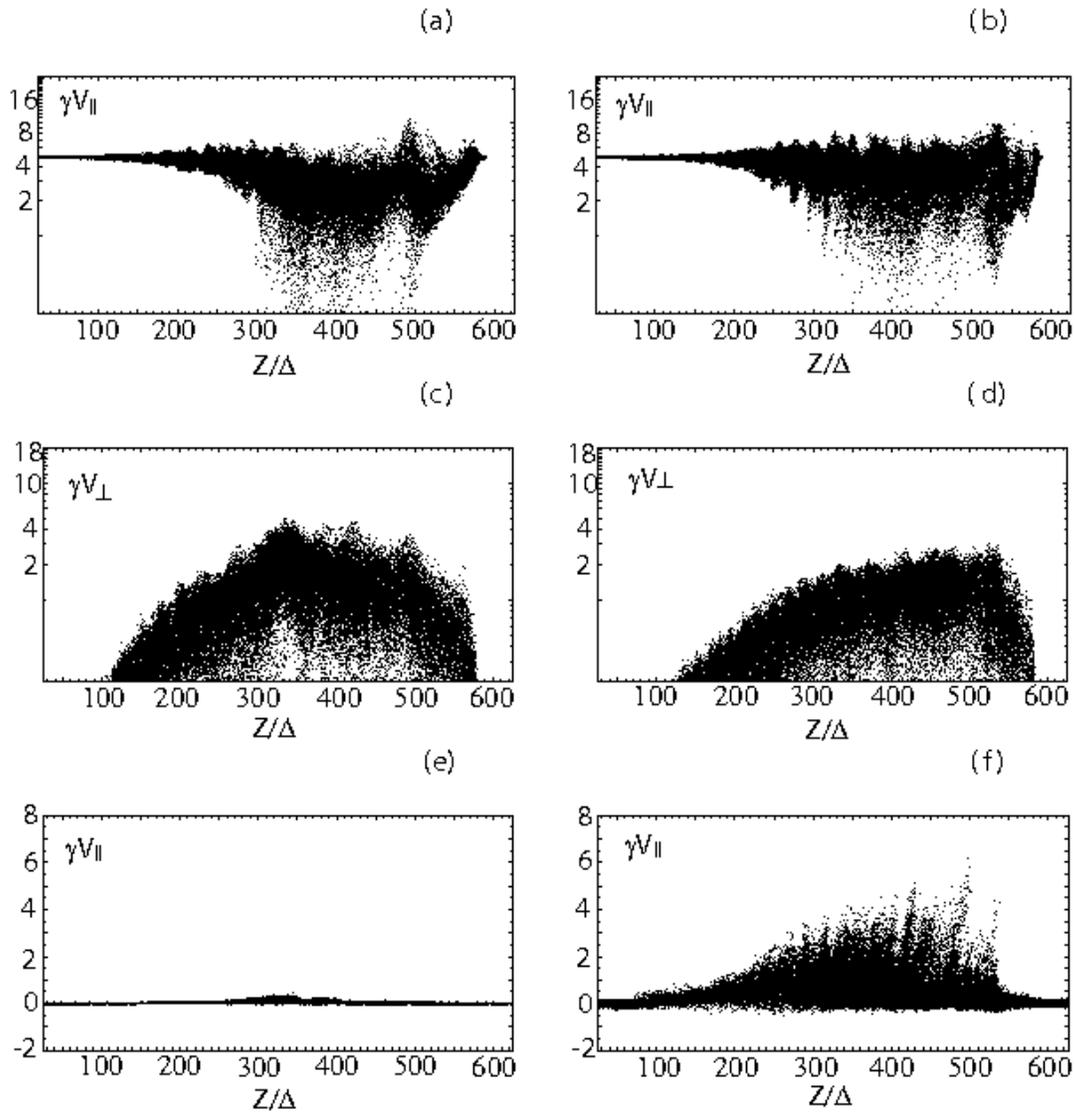}
\vspace*{0.5cm} \caption{Phase space distributions as a function of
$z$ are plotted for the  electron-ion (left column) and
electron-positron (right column) cases. Jet electrons in $Z/\Delta -
\gamma V_{\parallel}$ and  $Z/\Delta - \gamma v_{\perp}$ phase space
at $t = 59.8/\omega_{\rm pe}$ are plotted in top row and middle row,
respectively. In the bottom row the ambient electrons in $Z/\Delta -
\gamma V_{\parallel}$ phase space are plotted.}
\end{figure}

\clearpage

\begin{figure}[ht]
\epsscale{0.9} \vspace*{-2.0cm} \hspace*{-0.cm} \plotone{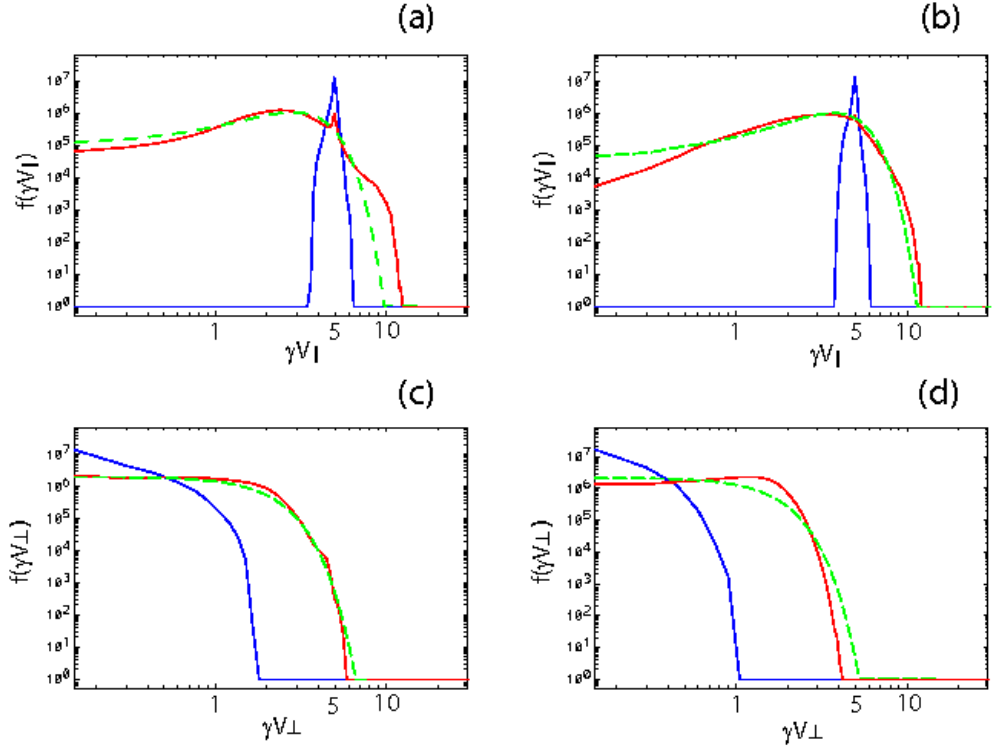}
\vspace*{-0.5cm} \caption{Jet electron parallel (panels a \& b) and
perpendicular (c \& d) momentum distributions are plotted at $t =
59.8/\omega_{\rm pe}$. The accelerated jet electron distribution is
shown by red curves and the initial distribution by blue curves for
the electron-ion (panels a \& c) and electron-positron (panels b \&
d) cases. The dashed green curves show a best fit Lorentz-boosted
thermal electron distribution in the parallel and perpendicular
directions. The zero count on the logarithmic scale is set to 1.}
\end{figure}

\end{document}